\documentclass[10pt,fullpage]{article}
\usepackage{fullpage}
\usepackage{cite}
\usepackage{url}
\usepackage{amsmath,amssymb,amsfonts}
\usepackage{algorithmicx}
\usepackage{algorithm}
\usepackage{algpseudocode}
\usepackage{graphicx}
\usepackage{physics}
\usepackage{textcomp}
\usepackage{hyperref}
\usepackage{xcolor} 
\usepackage{booktabs}
\def\BibTeX{{\rm B\kern-.05em{\sc i\kern-.025em b}\kern-.08em
    T\kern-.1667em\lower.7ex\hbox{E}\kern-.125emX}}

\title{MLQAOA: Graph Learning Accelerated Hybrid Quantum-Classical Multilevel QAOA}

\author{Bao Bach \\ 
\textit{Computer and Information Sciences} \\ 
\textit{Quantum Science and Engineering} \\ 
\textit{University of Delaware} \\ 
Newark DE, USA \\ 
{\tt baobach@udel.edu} \and 
 Jose Falla \\
 \textit{Physics and Astronomy} \\
 \textit{University of Delaware}\\
 Newark DE, USA \\
 {\tt jfalla@udel.edu} \and
 Ilya Safro \\
 \textit{Computer and Information Sciences} \\
\textit{Physics and Astronomy} \\
\textit{University of Delaware}\\
Newark DE, USA \\
{\tt isafro@udel.edu}
}
\begin{document}%\\

\maketitle

\begin{abstract}
Learning the problem structure at multiple levels of coarseness to inform the decomposition-based hybrid quantum-classical combinatorial optimization solvers is a promising approach to scaling up variational approaches. We introduce a multilevel algorithm reinforced with the spectral graph representation learning-based accelerator to tackle large-scale graph maximum cut instances and fused with several versions of the quantum approximate optimization algorithm (QAOA) and QAOA-inspired algorithms. The graph representation learning model utilizes the idea of QAOA variational parameters concentration and substantially improves the performance of QAOA. We demonstrate the potential of using multilevel QAOA and representation learning-based approaches on very large graphs by achieving high-quality solutions in a much faster time.\\
{\bf Reproducibility:} Our source code and results are available at \url{https://github.com/bachbao/MLQAOA}

%Combinatorial optimization is a crucial field with wide-range applications spanning engineering, logistics, telecommunications, and bioinformatics. Yet, the complexity of combinatorial problems can be formidable and influenced by various factors and especially, the scale of the problems.  Quantum approximate optimization algorithm (QAOA), leveraging the intrinsic principles of quantum mechancis and inspired by adiabatic evolution, is a promising candidate to tackle these problems. In this study, we delve into the Multilevel Scheme QAOA (MLQAOA) as a strategy to enhance the scalability of the QAOA algorithm on MAXCUT problem. By employing different quantum approaches as sub-problem solvers, we aim to address larger problem instances efficiently. Our investigation showcase the robustness and capabilities of MLQAOA across diverse graph settings, with biggest graph size up to 100,000 nodes. Moreover, experimental results underscore a comparable or even superior performance of MLQAOA when comparing to classical algorithms. This highlights the potential of MLQAOA to leverage limited quantum resources, particularly within the NISQ (Noisy Intermediate-Scale Quantum) era hardware.
\end{abstract}

\section{Introduction}

A general strategy for tackling many large-scale computational science problems on various hardware architectures, including fundamental combinatorial optimization problems on graphs, is the use of multilevel algorithms (also known as multiscale, multiresolution, and multigrid methods) \cite{brandt2003multigrid}. The multilevel approach starts by coarsening the problem to create a series (also known as a hierarchy) of progressively simpler, related problems at coarser levels, which are more feasible for the currently available hardware. This strategy is particularly useful in quantum computing where the number of qubits is limited as well as their overall quantum circuit fidelity, depth, and qubit connectivity. As a result, the hybrid quantum-classical algorithm developers put a lot of effort into making the circuit more compact. In particular, this is highly relevant to variational quantum algorithms \cite{Cerezo_2021} (e.g., many versions of the variational quantum eigensolver and quantum approximate optimization algorithm) when the same or slightly updated quantum circuit is reparametrized at each iteration and executed many times.

The motto of the multilevel algorithms is ``Think globally but act locally''. At each coarse level $i$, the best-found solution serves as an initialization for the next finer solution at level $i-1$. This initialization is enhanced through what is commonly referred to as ``local processing'' (also known as a refinement), a cost-effective series of fast steps that involve only a few variables at a time but collectively revisit all variables of that level multiple times. This method fits well with quantum computers, as it allows for solving parts of the problem within the constraints of limited qubit numbers.

In the classical domain, typical examples of such local processing include several iterations of classical relaxation methods like Gauss-Seidel or Jacobi relaxations when solving systems of equations, a few Monte Carlo passes in statistical physics simulations, or Kernighan-Lin or Fiduccia–Mattheyses search for optimization on graphs \cite{brandt2003multigrid}. In the quantum context, such multilevel framework was explored for the graph partitioning, clustering, and the maxcut problems. In both cases, the main local processing component was the quantum approximate optimization algorithm (QAOA). While these multilevel frameworks played their role and allowed increasing the instance graph size, the overall running time was suffering from the variational loop slowness. In this paper, we introduce graph representation learning-based hybrid quantum-classical multilevel QAOA and quantum-inspired recursive optimization frameworks that break the performance barrier and improve the overall solution quality. 

%Combinatorial optimization \cite{combinatorial} lies in the intersection of applied mathematics and theoretical computer science that focuses on finding the best solution from a finite discrete set of possible solutions. It deals with problems where the goal is to find an optimal arrangement or selection of elements, subject to certain constraints. However, as the size of combinatorial problems grows, so does their complexity, with the search space expanding exponentially. Consequently, huge efforts are devised to design fast algorithms for combinatorial optimization. Despite the ongoing efforts, the inherent complexity of the majority of combinatorial problems \cite{karp72} imposes a significant computational overhead on existing methods. Hence, there is a need for the development of more robust algorithms capable of efficiently accelerating the optimization process.

%As quantum computing continues to advance, researchers are exploring ways to leverage its unique properties to tackle combinatorial problems. Among those, Quantum approximate optimization algorithm (QAOA) 
The QAOA \cite{farhi2014QAOA} stands out as a promising candidate, offering the potential to achieve a speedup to certain combinatorial problems or classes of instances. 
%, QAOA demonstrates a superior approximation ratio relative to any known polynomial-time classical algorithm \cite{farhi2015quantum}. 
However, the current landscape of quantum computers, characterized by the NISQ (Noisy Intermediate-Scale Quantum) era, poses various challenges \cite{Bharti_2022}. Large-scale implementation of QAOA remains impractical \cite{guerreschi2019qaoa} due to physical limitations inherent in current quantum devices, such as constraints related to connectivity and noise. These hardware limitations present significant obstacles to realizing the full potential of QAOA in real-world applications. Taking into consideration the constraints and limitations of current quantum hardware, the hybrid architecture of QAOA emerges as one of the most promising approaches to mitigate the impact of hardware constraints by incorporating classical counterparts \cite{Cerezo_2021}. 

\noindent {\bf Our contribution} We introduce a scalable hybrid quantum-classical multilevel scheme integrated with two QAOA-inspired solvers and accelerated by the graph representation learning for fast parametrization of QAOA \cite{jose2024graph}. This scheme allows scalability for fast small-scale QAOA solvers by decomposing the original problem across the different scales of coarseness into sub-problems and constructing the global solution from the sub-problem solutions. This resolves the huge time complexity that was a significant barrier in previous hybrid quantum-classical decomposition-based algorithms \cite{shaydulin2019hybrid,ushijima2021multilevel,angone2023hybrid} and makes them comparable even to purely classical algorithms. 

Specifically, our work focuses on addressing the maximum cut (MAXCUT) problem on large-scale graphs using the graph representation learning-based parameter transferability for QAOA \cite{jose2024graph} and quantum-informed recursive optimization \cite{finzgar2024quantuminformed} as accelerators. These methods enhance the execution time of the sub-problem solver, resulting in an overall speed-up of the multilevel scheme with two orders of magnitude compared with \cite{shaydulin2019hybrid,ushijima2021multilevel,angone2023hybrid} approaches on mid-scale graphs and \emph{way more on larger that (to the best of our knowledge) no hybrid quantum-classical approach was able to tackle in a reasonable computational time}. Our approach not only yields faster runtimes but also improves the performance of multilevel QAOA (it was impossible to run other QAOA-based methods due to the slow performance and/or huge memory requirements). To validate our approach, we evaluate it on diverse sets of graphs, including real-world problems like social networks, optimization instances, graphs that are hard for MAXCUT, and those that are particularly hard for the Goemans-Williamson MAXCUT approximation\cite{yyyeGset, karloff_graphs,davis2011university, Rossi_Ahmed_2015}. Even in the QAOA simulation mode, our experimental results demonstrate competitive quality compared to the MAXCUT dedicated state-of-the-art classical solvers, with comparable runtime.
\section{Preliminaries and Notations}
\subsection{QAOA and MAXCUT}
\label{sec:qaoa_maxcut}
Quantum approximate optimization algorithm (QAOA) \cite{farhi2014QAOA} is a hybrid quantum algorithm focusing on combinatorial optimization problems. The quantum heuristic algorithm aims to produce an approximation of the problem's solution by alternating between cost-function-based Hamiltonian and mixing Hamiltonian. The QAOA variational loop consists of $p$ parameterized layers of alternating unitary operators and a classical optimizer. The role of the optimizer is to find the best set of parameters that minimize the cost function of the problem. 

Given a combinatorial optimization problem with a cost function $f(x)$ where $x \in \{0, 1\}^{n}$, QAOA alternatingly apply unitaries drawn from two Hamiltonian families, cost-function-based unitary $U_{P}(\gamma) = e^{-i\gamma H_f}$ and mixing unitary $U_{M}(\beta) = e^{-i\beta H_B}$ parametrized by $\gamma = \{\gamma_i\}$ and $\beta = \{\beta_i\}$, $1\leq i \leq p$, respectively. Hamiltonian $H_f$ is a cost function-based Hamiltonian where the information of cost function $f(x)$ is embedded while Hamiltonian $H_B$ is a fixed mixing Hamiltonian. With $H_f$ as the observable, QAOA prepares the quantum state expressed in (\ref{eq: QAOA_state}) and performs optimization concerning the expectation value $\langle H_f \rangle = \bra{\gamma,\beta}H_f\ket{\gamma,\beta}$.
\begin{equation}
    \label{eq: QAOA_state}
    \ket{\gamma,\beta} = U_{M}(\beta_p)U_{P}(\gamma_p)\dots U_{M}(\beta_1)U_{P}(\gamma_1)\ket{+}^{\otimes n}
\end{equation}
The MAXCUT problem on a graph is the first QAOA demonstration \cite{farhi2014QAOA} and the algorithm's most common benchmark. This problem is NP-complete  \cite{NP_completeQAOA}. The problem involves finding a cut that splits the graph nodes into two disjoint parts $V_1$ and $V_2$ such that the weighted sum of edges $ij$  connecting two parts is maximized. The MaxCut problem is often formulated as a quadratic unconstrained binary optimization problem (QUBO) by assigning a binary value $x_i$ to every node based on its part. Given a graph $G(V, E, w)$ where $V$ is the set of nodes, $E$ is the set of edges and $w$ is the edge weighting function, the MAXCUT problem is defined in (\ref{eq: MaxCut_objective}). 
\begin{equation}
    \label{eq: MaxCut_objective}
    \max_{x\in \{-1, 1\}^n} \sum_{ij \in E} w_{ij}\frac{(1-x_{i}x_{j})}{2}
\end{equation}
For this QUBO problem, the Hamiltonian can be constructed by applying the mapping $x_i \mapsto \frac{1}{2}(1-Z_i)$, where $Z$ is the Pauli operator $Z$.

\subsection{Multilevel Methods}

The multilevel methods for optimization on graphs are inspired by the multigrid methods that were originally devised to tackle boundary value problems in spatial domains \cite{brandt2003multigrid}. Choosing a set of grids makes these problems discrete and consists of algebraic equations associated with the grid points. The essence of the multigrid method lies in iteratively increasing the grid point spacing, transforming the original problem into coarser versions, and leveraging solutions from these coarser problems to aid in finding the final solution. The theory behind this scheme is based on two observations. First, the standard iterative methods have smoothing properties, this means they are effective at relaxing oscillatory components of error while leaving smooth components unchanged. Second, the smooth modes on a fine grid look less smooth on a coarse grid. Those observations imply that incorporating coarse grids during computation can make the smooth components of error of the finest grid look more oscillatory and be eliminated by the relaxation of the iterative method. The idea of such elimination and some ways of constructing the coarse problems were inherited from the multilevel methods.

Nowadays, multigrid-inspired multilevel methods are applied to a broader spectrum of problems regardless of their connection with physical grids. A broader organizational framework has emerged, replacing the concept of a coarse grid with a more generalized notion known as the multilevel method \cite{brandt2003multigrid}. The approach proved itself to be useful in the (hyper)graph context where the multilevel method utilizes the graph structure to curtail large-scale graphs into their coarser representations. Examples of problems successfully tackled by multilevel methods include graph partitioning \cite{safro2012advanced}, visualization \cite{hu2015visualizing}, linear ordering \cite{safro2006graph}, and generation \cite{chauhan2019multiscale}. 

%The multilevel method yields the same principle as multigrid, by solving coarsen problems with high quality and leveraging that to refine the final solution at the finest stage. 
While there are several types of the coarsening-uncoarsening schedules in multilevel methods (e.g., V-, W-, or FMG cycles \cite{brandt2003multigrid}), in this work we use the simplest $V-$cycle to eliminate additional advantages of the classical computing in the hybrid framework. In this setting, we generate a hierarchy of next coarser graphs 
\begin{equation}\label{eq:gi}
\{G_l=(V_l, E_l, w_l)\}_{l=0}^L,
\end{equation}
where $l$ is the index of level, $G_0$ is the original large-scale graph, and $G_L$ is the coarsest smallest graph. The coarsening consists of (1) relaxation-based grouping pairs of nodes based on the recently introduced maximization version \cite{angone2023hybrid} of the algebraic distance for graph coarsening \cite{chen2011algebraic}, and (2) edge weight accumulation.

After the hierarchy is created, the MAXCUT at the coarsest level is solved and the solution is gradually uncoarsened all the way up to the finest level. At each step of the uncoarsening, the $l$th level solution is initialized by that from level $l+1$ and further refined by various QAOA approaches. 
%With original graph $G_0(V_0, E_0)$ where $V$ is the set of vertices and $E$ is the set of edges, the coarsening stage creates a sequence of $N$ increasingly coarser graphs $G_1(V_1, E_1), G_2(V_2, E_2), \dots, G_N(V_N, E_N)$ where $\abs{V_1} > \abs{V_2} > \dots > \abs{V_N}$ and $\abs{E_1} > \abs{E_2} > \dots > \abs{E_N}$, the notation $G_{k}$ address the graph at level $k$ where $0 \geq k \leq N$. Usually, this phase includes grouping nodes and accumulating edge weights. For the next phase, the solver will find a high-quality solution to the problem for the coarsest graph $G_n$. From this solution, an iterative process happens, going from the coarsest to the finest level. At each level, using the last level solution as the initialization, the uncoarsening phase occurs by separating the merged nodes at that level to get from coarse to fine level. After the initialization, random local sub-problems are generated and solved to refine the current solution. 
In the end, the final solution of the original problem is obtained.

%\subsection{Graph Learning For Parameter Transferability}
\subsection{Graph Representation Learning}
Graph representation learning techniques have shown great promise in addressing graph analytic tasks, such as node classification, link prediction, and community detection by transforming graph features (nodes, edges, edge weights, etc.) into non-linear, low-dimensional, dense, continuous, and highly informative vector spaces \cite{cai_comprehensive_2018}. In these low-dimensional graph representations, if two graph instances possess common structural features, they exhibit closeness with respect to some distance function, such as a Euclidean distance function. In this work, this idea is utilized for the scalable transferability of QAOA parameters.

Of these graph representation techniques, some involve node-level embedding %\cite{grover_node2vec_2016} %, sybrandt2020fobe, ding2020unsupervised, rozemberczki_fast_2020, yang_nodesketch_2019} 
at various scales (microscopic, mesoscopic, and macroscopic node embedding), while others (that are relevant to this work) involve whole graph embedding. Whole graph embedding techniques, as previously mentioned, are useful when analyzing whole networks \cite{wang_graph_2021, cai_simple_2022, galland_invariant_2019, narayanan_graph2vec_2017}, particularly when trying to determine the structural similarity between graphs.

Methods for graph representation learning can be grouped into a few categories. Among these categories, a classic family of methods involves graph kernels, with examples including the Weisfeiler-Lehman \cite{shervashidze_weisfeiler-lehman_2011}, random walk \cite{gartner_kernels_2003}, shortest path \cite{borgwardt_shortest-path_2005}, and deep graph \cite{yanardag_deep_2015} kernels. Another family of methods involves graph embedding for learning vector representation of graphs, with examples including Graph2Vec \cite{narayanan_graph2vec_2017}, which uses the Weisfeiler-Lehman kernel to extract rooted subgraph features to obtain the embeddings; also, GL2Vec \cite{chen_gl2vec_2019}, which is an extension of Graph2Vec that includes line graphs to account for edge features. 
%The Geo-Scatter \cite{gao_geometric_2019} and FEATHER \cite{rozemberczki_characteristic_2020} methods employ normalized adjacency matrices to capture the probability distribution of neighborhoods in graphs. 
More recently, an approach to graph representation learning involves graph neural networks, which employ machine learning methods, with some examples including GCN \cite{welling_semi_2016}, SGCN \cite{weinberger_simplifying_2019}, GIN \cite{jegelka_gnns_2018}, and Causal GraphSAGE \cite{zhang_causal_2022}, to name a few. Finally, there is a family of methods that use information from the graph's Laplacian matrix and its eigenvalues to generate embeddings, such as SF \cite{de_lara_simple_2018}, NetLSD \cite{tsitsulin_netlsd_2018}, and FGSD \cite{verma_NIPS_2017}. Our method belongs to the latter type of method.
\section{Related works}

\subsection{QAOA-based Approaches}

At the heart of this algorithm lies the utilization of QAOA to optimize sub-problem graphs, which refines the final solution on the finest graph. Hence, the quality of the QAOA solution serves as a benchmark for our refinement performance. In this study, we employ vanilla QAOA with three layers, bypassing the exhaustive optimization process by incorporating graph learning techniques for parameter transferability \cite{jose2024graph, galda2023similarity}. This approach is one of many attempts to enhance the performance of QAOA, some noticeable attempts include using warm-start for good initialization \cite{Egger_2021}, noise-directed adaptive mapping for a higher-quality solution based on previous step result  \cite{maciejewski2024improving}, finding the best QAOA ansatz architecture for the given Hamiltonian with bayesian optimization \cite{duong2022quantum} and the connection between underlying symmetry of the objective function and QAOA \cite{Shaydulin2021}.

\subsection{Hybrid Quantum-Classical Algorithms For Large Graphs}
Quantum algorithms like the Quantum Approximate Optimization Algorithm (QAOA) hold significant promise in addressing complex graph problems. However, their efficacy is hindered by the current limitations of quantum hardware, which often suffer from noise and architectural constraints. Consequently, to overcome these limitations and scale up the application of quantum algorithms such as QAOA to tackle larger graph instances, researchers explore strategies to downscale the original problems into manageable subproblems. The first attempts to utilize multilevel methods in a hybrid quantum-classical setting,  \cite{angone2023hybrid,ushijima2021multilevel} introduced the coarsening-uncoarsening approach and solved multiple small graphs at each level to enhance the final solution of the original larger graph problem. This work extended the single-level decomposition-based approaches \cite{shaydulin2019hybrid}. However, already with the subproblem graph size of 22 nodes, the solvers were too slow due to the parametrization in variational loops. This scalability issue is addressed in this work. 

In a similar vein, \cite{ponce2023graph} employs the graph decomposition technique to scale down the original graph to a manageable size, producing a high-quality approximation of the final solution from the simplified problem. In addition to the down-scaling approach, the divide-and-conquer method stands out as another candidate. As exemplified by \cite{QAOA-in-QAOA}, the work demonstrates the potential of tackling divided subgraphs in parallel and then merging these solutions to derive the final solution. This approach is comparable to the two-level scheme if viewed as a multilevel algorithm.

From the circuit cutting, \cite{Tang_qccut} creates a hybrid scalable approach CutQC to distributing big quantum circuits that can be run on smaller quantum processing units (QPU), this allows breaking the limit of classical simulation and achieving a larger quantum circuit evaluation. Similarly, \cite{Smith_qccut} introduces SuperSim which uses Clifford-based circuit cutting. By isolating (cutting) the Clifford circuit within the big non-clifford circuit, the approach leads to the utilization of Clifford simulation that greatly reduces runtime. %Another promising approach  \cite{keller2023hierarchical} is  inspired by multigrid but used as an alternative to QAOA with the genral goal of variational parameter repurposing. 

% An intriguing approach in this regard involves leveraging quantum information from the QAOA solution to recursively simplify the graph, as proposed in \cite{finzgar2024quantuminformed}.

\subsection{Graph Learning For Parameter Transferability}
The task of finding good QAOA parameters is challenging in general. For example, determining whether the optimized solution corresponds to a local or a global minimum in the energy landscape, or due to encountering barren plateaus \cite{anschuetz_beyond_2022, wang_noise-induced_2021}. Furthermore, while the QAOA solution improves as the depth of the circuit is increased, it only does so marginally at the cost of increasing the computational complexity of optimizing the variational parameters \cite{shaydulin_2019_eval}. For this reason, acceleration of optimal parameter search for a given QAOA depth \textit{p} is a key in demonstrating quantum advantage. Examples of optimal parameter search acceleration include warm- and multi-start optimization \cite{Egger_2021, shaydulin2019multistart}, problem decomposition \cite{shaydulin2019hybrid}, instance structure analysis \cite{Shaydulin2021}, and multigrid inspired parameter learning \cite{keller2023hierarchical}. In particular, optimal QAOA parameter transferability has shown great promise in circumventing the problem of finding good QAOA parameters \cite{galda2021transferability, galda2023similarity}. Based on structural graph features, successful parameter transferability can be achieved between a donor graph and an acceptor graph, with a very small reduction in the approximation ratio.
\section{Our methods}

\begin{figure*}[!htb]
    \centering
    \includegraphics[width=1.0\textwidth, height = 0.475\textwidth]{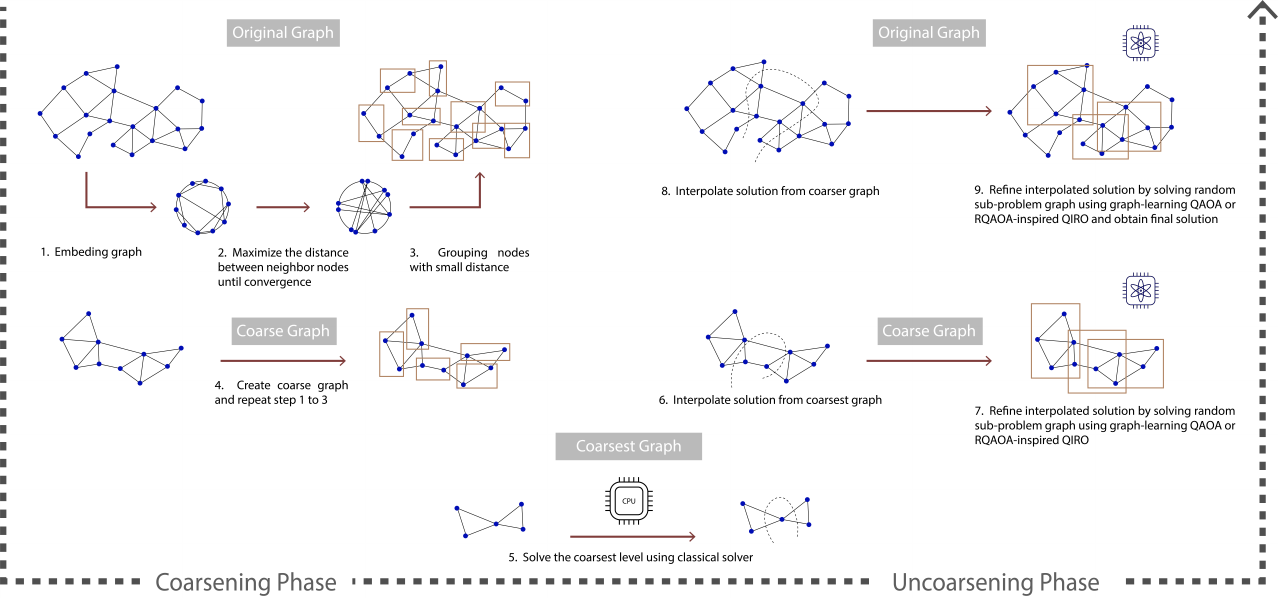}
    \caption{MLQAOA scheme using $V$-cycle. The original graph is iteratively coarsened in the Coarsening Phase. The coarsest graph is solved using a classical solver. Building upon this solution, the Uncoarsening Phase leverages the previous level solution through interpolation and performs refinement by solving sub-problem graphs.}
    \label{fig:overview}
\end{figure*}

The multilevel MAXCUT QAOA approach, as introduced in \cite{angone2023hybrid}, encounters a significant challenge stemming from the substantial computational overhead incurred in refining numerous sub-problem graphs by slow variational loops. In some cases, this issue leads to inefficiencies even when compared to general-purpose global solvers such as Gurobi \cite{gurobi}. To address this limitation, we leverage the potential of graph learning techniques for QAOA parameter transferability, as outlined in \cite{jose2024graph} and the use of single and second correlations to simplify the sub-problem graph \cite{rqaoa}. Our proposed method (MLQAOA) introduces a reinforcement for this multilevel scheme. By using a graph learning model tailored to the weighted graphs at each level or quantum-informed recursive optimization, the model then effectively approximates high-quality ansatz parameters, generating high-quality solutions for the sub-problems. Figure \ref{fig:overview} demonstrates  our method.  

\subsection{Coarsening Phase}
At each level of coarseness (i.e., at each $G_l$ in (\ref{eq:gi})), the coarsening phase is started by constructing a $d$-dimensional unit sphere with every node from the graph $G_l$ embedded and further relaxed in. This is essential for establishing a distance metric between nodes for multilevel algorithms, as discussed in \cite{ron2010relaxationbased}. Every node is initially placed in a random position on the surface of the sphere or $d$-dimensional hypercube embedded in the sphere. Denote $p^{t}_{i}$ as the position of node $i$ at iteration $t$. For example, for the initial hypercube embedding $\forall {i} \in V_l, p^{0}_{i} \gets \text{rand} [ -1, 1]^{d} $.

Following the initialization, several node-wise correction iterations are applied to maximize within the sphere the total weighted distance between each node and its neighbors:  
\begin{equation}\label{eq:relax}
    \forall {i} \in V_l, p^{t+1}_{i} \gets \max{\sum_{j \in N(i)}} w_{ij} \norm{p^{t}_{i} - p^{t}_{j}}_{2}.
\end{equation}
Here, $N(i)$ represents the set of neighbors of node $i$, and $w_{ij}$ is the weight of the edge $ij$.  This iterative scheme continues until convergence is achieved. 

Subsequently, by leveraging the embedding, the nodes are paired for further coarsening using a K-D tree. A fast search within the K-D tree facilitates pairing the node with its nearest unpaired neighbor. This search finalizes the formation of the new coarse graph by contracting every paired node in the matching and creating the coase nodes for $V_{l+1}$. Algorithm \ref{alg:coarsening} describes the coarsening scheme. This process is repeated until the desired coarsest graph size is achieved. Finally, by grouping into pairs of nodes at each level, the original graph is broken into approximately $\log |V_0|$ number of increasingly coarse graphs.
\begin{algorithm}
    \caption{Coarsening}
    \label{alg:coarsening}
    \begin{algorithmic}[1]
        \Require \text{Graph at level }$l$,   $G_{l} = (V_{l}, E_{l}, w_{l})$
        \State $A_{l} \gets$ Adjacency matrix of $G_{l}$
        \State $\text{Initialize embedding of } G_{l}$ \text{ into sphere}
        \State $\mathfrak{E} \gets \text{ Apply iterations of (\ref{eq:relax})}$ \Comment{$V_l$ is embedded into sphere}
        
        \State $matched \gets \emptyset$; $pairs \gets \emptyset$
        \For{$i \in V_{l}$} \Comment{Pair node with nearest neighbor in $\mathfrak{E}$}
            \If{$i \notin matched$}
                \State $j \gets \mathfrak{E}\text{-nearest not matched neighbor of }i$
                \State $matched \gets matched \cup \{i, j\}$
                \State $pairs \gets pairs \cup (i, j)$

            \EndIf
        \EndFor
        \State \Comment{Now each pair is a node in $V_{l+1}$ and $|V_{l+1}| = |pairs|$}
        \State $P \gets 0^{\abs{V_{l+1}} \times \abs{V_l}}$ \Comment{Initialize multilevel restriction operator. For details see \cite{ron2010relaxationbased}.}
        \State $q \gets 0$ \Comment{Initialize iterator over coarse nodes}
        \For{$(i,j) \in pairs$} \Comment{Construct coarse graph}
            \State $P_{q,i} \gets 1$; $P_{q,j} \gets 1$
            \State $q \gets q + 1$
        \EndFor
        \State $A_{l+1} \gets P^{T}A_{l}P$ \Comment{This is the algebraic multigrid formulation to create the coarse graph with weighted edges that are accumulated from the fine level. In general, algebraic multigrid is not restricted to having only two fine nodes to form a coarse node. For details see \cite{ron2010relaxationbased}.}\\
        \Return $G_{l+1}$ obtained from $A_{l+1}$
    \end{algorithmic}
\end{algorithm}

\subsection{Uncoarsening Phase}
The uncoarsening phase starts with the coarsest graph $G_L$ obtained in the last coarsening step. The number of nodes in $G_L$ is based on the sub-problem size that is defined by the user. The solution to the problem of the coarsest graph is found through state-of-the-art classical solver MQLib employing the BURER2002 heuristic \cite{DunningEtAl2018} or extensively optimized QAOA. Usually, the larger $|V_L|$ could be, the better it affects the final results.

Building upon the initial solution of $G_L$, the uncoarsening phase progresses towards a finer level through linear interpolation from the previous coarser level solution. This interpolation is a surjective mapping $F: V_{l+1} \rightarrow V_{l}$ (i.e., mapping of a fine node to its aggregating coarse node), and the initial solution at the fine level is initialized by  $\forall i \in V_{l+1}, x_{i} = x_{F(i)}$. %The mapping aligns vertices from finer graph $V_{f}$ to vertices of the coarser graph $V_{c}$.
This %At level $l$, the interpolation of the solution at level $l+1$ 
yields an initial approximation for further refinement. 

The refinement scheme begins with the computation of the gain of every node. This gain is then used to estimate the impact of each node on the final energy at the current level and update the MAXCUT objective. This gain is efficiently tracked by updating it based on the edges that enter or leave the cut.   
\begin{equation}\label{eq:gain}
    \forall i \in V_{l}, gain(i) \gets \sum_{j \in N(i)} w_{ij}(-1)^{2x_{i}x_{j} -x_{i}-x_{j}}
\end{equation}
During the refinement stage, sub-problems are iteratively generated and solved to improve the final solution. In each iteration, $n$ random nodes are sampled and ranked by their gains, and $K$ nodes with the highest gain are then used to construct the sub-problem graph. This construction involves creating a graph with two super-nodes, along with the selected nodes. Each super-node aggregates the nodes not chosen for the sub-problem, and weighted edges are added between the super-nodes and chosen nodes.

Given the sub-problem graph, a sub-problem graph solver scheme must be devised. In this work, instead of using exhaustive learning on vanilla QAOA, we study how the graph learning for parameter transferability for QAOA with depth $p=3$ \cite{jose2024graph} and quantum-informed recursive optimization algorithm \cite{finzgar2024quantuminformed} works as a sub-problem solver. This investigation follows the spirit of the classical multilevel algorithm. By leveraging an excellent efficient, small-scale problem solver, the multilevel approach enables the scalability of the problem solver with good approximation. %\color{red}
%Should write more in here to explain the reason why we are doing this -- Bao, Ilya: No
%\color{black}

After the relaxation, if the sub-problem solution enhances the objective function, the solution and objective are updated. The iteration counter is reset whenever there is an improvement. To prevent excessive iterations, a restriction is set: if three consecutive iterations do not yield improvement or if there have been a total of ten iterations, the algorithm accepts the current solution and proceeds to the next level. Algorithm \ref{algo:refinement} outlines the process.
\begin{algorithm}
    \caption{Refinement}
    \label{algo:refinement}
    \begin{algorithmic}[1]
        \Require Graph $G_{l}(V_{l},E_{l})$,  Subproblem graph size $K$, Initial solution from previous level $S_{l+1}$
        \State $\mathfrak{G} \gets$ compute gain for all nodes as in Eq. (\ref{eq:gain}) 
        \State $maxIter, count \gets 0$
        \State $\mathcal{O} \gets$ compute objective inherited from $S_{l+1}$
        \While{$count < 3$ and $maxIter < 10$}
            \If{$count = 0$}
                \State $n \gets \abs{V_{l}}$
            \Else
                \State $n \gets max(0.3\abs{V_{l}}, 2K)$
            \EndIf
            \State $subset \gets K$ highest gain nodes from randomly sampled $n$ nodes from $V_{l}$ 
            %\State $subproblem \gets K$ highest gain nodes from $subset$
            \State $P\gets$ constructMAXCUTSubproblem($subset$)
            \State Solve $P$ and compute new objective $\mathcal{O}^{new}$
            %\State $solveSubproblem()$
            %\State $newSolution \gets updateSolution$
            %\State $newObjective \gets Objective(newSolution)$
            %\State $updateGain()$
            \State Update $\mathfrak{G}$
            \State $maxIter \gets maxIter + 1$
            \If{$\mathcal{O}^{new} \geq  \mathcal{O}$}
                \State $\mathcal{O} \gets \mathcal{O}^{new}$
                \State $S_{l} \gets $ new solution derived from $\mathcal{O}^{new}$
                \State $count \gets 0$
            \Else
                \State $count \gets count + 1$
            \EndIf
        \EndWhile
        \State \Return $S_{l}$
    \end{algorithmic}
\end{algorithm}

\subsection{Graph Representation Using Spectral Learning For Parameter Transferability On Weighted Graphs}
We begin the graph representation by defining the Laplacian matrix:
\begin{equation}
    L = D - A,
\end{equation}
where $D$ is the degree matrix and $A$ the adjacency matrix. Denoting the vector $w = (w_{1}, ..., w_{n})$ as the weights for each node $(1, ..., n)$, $D = \text{diag}(d)$, where $d = Aw$. The Laplacian is a symmetric, positive semi-definite matrix. Observing that,
\begin{equation}
    \forall v \in \mathbb{R}^{n}, \quad v^{T}Lv = \sum_{i<j}A_{ij}(v_{j} - v_{i})^{2}
\end{equation}
the spectral theorem yields:
\begin{equation}
    L  = U\Lambda U^{T},
\end{equation}
where $\Lambda = \text{diag}(\lambda_{1}, ..., \lambda_{n})$ is the diagonal matrix of eigenvalues of $L$, with $0 = \lambda_{1} < \lambda_{2} \leq ... \leq \lambda_{n}$, and $U = (u_{1}, ..., u_{n})$ is the matrix of corresponding eigenvectors, with $U^{T}U = I$.

For our graph representation model, we consider the following normalized version of the Laplacian, referred to as the \textit{weighted} Laplacian:
\begin{equation}
    L_{W} = W^{-\frac{1}{2}}LW^{\frac{1}{2}},
\end{equation}
where $W = \text{diag}(w)$. Once again, this is a symmetric, positive semi-definite matrix, and the spectral theorem yields:
\begin{equation}
\label{eq:weighted_laplacian}
    L_{W} = \hat{U}\hat{\Lambda}\hat{U}^{T},
\end{equation}
where $\hat{\Lambda} = \text{diag}(\hat{\lambda}_{1}, ..., \hat{\lambda}_{n})$ is the diagonal matrix of the eigenvalues of $L_{W}$, with $0 = \hat{\lambda}_{1} < \hat{\lambda}_{2} \leq ... \leq \hat{\lambda}_{n}$, and $\hat{U} = (\hat{u}_{1}, ..., \hat{u}_{n})$ is the matrix of corresponding eigenvectors, with $\hat{U}^{T}\hat{U} = I$.

To determine the similarity between two graph instances, we compute the Euclidean distance between the first non-trivial $\hat{u}_{2}$ (second) eigenvector\footnote{Since we are working with connected graphs, the first eigenvector of the Laplacian is trivial.} of the weighted Laplacian matrix \ref{eq:weighted_laplacian} of each of the graphs. The reason for this is that the first non-trivial eigenvector of a graph's Laplacian contains information about the graph's connectivity, and therefore, its structure. We are interested in determining graph similarity based on structure since our previous work has shown that structural similarity is a key feature in determining successful parameter transferability between two graph instances \cite{galda2023similarity}.

The construction of the model for parameter transferability begins by generating a corpus of 22-node graphs. This corpus of graphs contains $\sim 5,000$ various types of graphs, including random graphs with weight distributions ranging from $w \in [0, m * 5)$ with $m \in (1, 2, 3, 4, 5, 6, 7, 8, 9, 10)$. Additionally, the corpus includes graphs that are structurally similar to those found in our multi-level approach. Each of the corpus 'graphs' optimal parameters is optimized on 20 independent multi-starts. For the 20 multi-starts, the best solution to the expectation value of the cost Hamiltonian $\langle H_f \rangle = \bra{\gamma,\beta}H_f\ket{\gamma,\beta}$ is taken as the graph's optimal solution, and the optimal parameters $(\Vec{\gamma}, \Vec{\beta})$ are stored\footnote{For a depth of $p = 3$, there are 3 optimal $\gamma$ and 3 optimal $\beta$ parameters.}. The circuit used for graph optimization is constructed as per Section \ref{sec:qaoa_maxcut} using Cirq \cite{cirq_developers_2023_10247207}, and the quantum-classical optimization loop is performed with a COBYLA solver for 300 iterations, or until convergence. In the end, the model consists of an $N \cross n$ matrix, where $N$ is the number of graphs in the corpus, and $n$ is the number of nodes, with each row corresponding to the first non-trivial eigenvector of each graph in the corpus.

\subsection{Quantum-Informed Recursive Optimization Algorithm}
QAOA is a $local$ algorithm, a characteristic that has been demonstrated to constrain its overall performance. To address this problem, recursive QAOA \cite{rqaoa} introduced non-local updates by iteratively eliminating variables using the single and double correlation information between them. This iterative process systematically prunes variables from the original optimization problem while creating new connections between previously distant pairs of nodes. The change in graph structure introduces a non-local effect that counteracts the inherent locality of QAOA. Based on this foundation, Quantum-Informed Recursive Optimization Algorithm (QIRO) \cite{finzgar2024quantuminformed} emerges as a family of approaches that leverages quantum information to derive the potential classical problem-specific reduction, thus recursively simplifying the original problem. 

The framework of QIRO is shown in algorithm \ref{QIRO}, where the original problem undergoes successive simplification until reaching a desired size. This reduction process exploits single and double correlations concerning a low-energy quantum state (lines 6, 7) and a problem-specific simplification rule (line 9). In this study, we adopt the simplification rule for the MAXCUT problem from \cite{rqaoa}, where the rule is defined with the use of correlation matrix $M$. Initially, we find the edge $(i, j) \in E_{it}$ exhibiting the largest magnitude of $M_{ij}$. Subsequently, we treat $\hat{Z}_{i}$ and $\hat{Z}_{j}$ as correlated if $M_{ij} > 0$ and anti-correlated if  $M_{ij} < 0$. By assigning $\hat{Z}_{j} = \text{sgn}(M_{i,j})Z_{i}$, we effectively eliminate variable $Z_{j}$ from our Hamiltonian, resulting a new Hamiltonian depending on only $\abs{V} - 1$ variable or a graph $G$ with $\abs{V} - 1$ nodes. We remark on the possibility that there may exist a better simplification strategy for the MAXCUT problem that we are not aware of.

To ensure RQAOA-inspired QIRO maintains a competitive runtime, we employ QAOA with a single layer, as the single and double correlation can be efficiently calculated \cite{Ozaeta_2022}.

\begin{algorithm}
    \caption{General scheme of QIRO algorithm}
    \label{QIRO}
    \begin{algorithmic}[1]
        \Require Graph $G_{0}(V_{0},E_{0})$
        \Require Smallest problem size $s$
        \State $l \gets 0$
        \State $S \gets \emptyset$
        \While{$\abs{V_{l}} > s$}
            \State Prepare low-energy quantum state $\ket{\psi}$
            \State $M \gets 0^{\abs{V_{l}} \times \abs{V_{l}}}$
            \State $\forall i \in V_{l}$, $M_{ii} \gets \bra{\psi}\hat{Z}_{i}\ket{\psi}$
            \State $\forall (i, j) \in E_{l}$, $M_{ij} \gets \bra{\psi}\hat{Z}_{i}\hat{Z}_{j}\ket{\psi}$
            \State $l \gets l + 1$
            \State $G_{l}(V_{l}, E_{l}), S \gets $ Simplification($G_{l-1}$, $M$, $S$)
        \EndWhile
        \State $S \gets  Classical\_Solver(G_{l})$
        \State \Return $S$
    \end{algorithmic}
\end{algorithm}

\section{Computational Results}
Our source code and results are available at \url{https://github.com/bachbao/MLQAOA}. To evaluate the performance and scalability of MLQAOA alongside the proposed sub-problem graph solvers, we conducted numerical simulations using large instance graphs sourced from diverse datasets. We present a detailed outline of our experimental setup, including hyperparameter configurations for MLQAOA and both sub-problem solvers, namely, Graph Learning Parameter Transfer and RQAOA-inspired QIRO. Subsequently, we offer a comprehensive comparison between our approach and classical solvers, focusing on approximation ratio and runtime metrics.

The configuration for MLQAOA sub-problem solvers is as follows: The sub-problem graph size, denoted by $K$, is set to 20. For parameter transferring using graph learning on weighted graphs, the QAOA entails a 22-qubit circuit, comprising 20 chosen nodes and 2 supernodes, with 3 layers. For RQAOA-inspired QIRO, the smallest size problem is 10. The simulations are executed on the ibmq\_qasm\_simulator  with 10240 sampling shots. 

We simulate our method on three different graph instance sets: (1) the well-known public MAXCUT dataset $G_{set}$ \cite{yyyeGset}, (2) the Karloff graphs \cite{karloff_graphs} and (3) larger graphs sourced from SuiteSparse Matrix Collection \cite{davis2011university} and the Network Repository \cite{Rossi_Ahmed_2015}. The selection of these graphs is based on their widespread use in MAXCUT problems, their diversity in size, degree distributions, and edge weights, and their challenging nature, as demonstrated by their effectiveness as benchmarks for existing solvers. Especially, the Karloff collection contains challenging graphs for the Goemans-Williamson MAXCUT algorithm \cite{GW_algorithm}, the polynomial-time approximation algorithm.% for MAXCUT with the best-known approximation ratio.

In all results,  we use either the approximation ratio (best known or optimal solution vs those by our algorithms) or the best objective cut as performance metrics to compare our proposed scheme with other solvers. Specifically, given graph $G$ and MAXCUT problem, the best objective cut achieved by method $\mathcal{A}$ is denoted as $C_{\mathcal{A}}$ while the optimal cut (if known) is denoted as $C^{*}$. The approximation ratio of method $\mathcal{A}$ is defined as $r_{\mathcal{A}} = \frac{C_{\mathcal{A}}}{C^{*}}$. As finding the optimal cut for large graphs is not always possible, we primarily use the approximation ratio metric for graphs with known optimal cuts.

\begin{figure*}[!htb]
    \centering
    \includegraphics[width=0.8\textwidth]{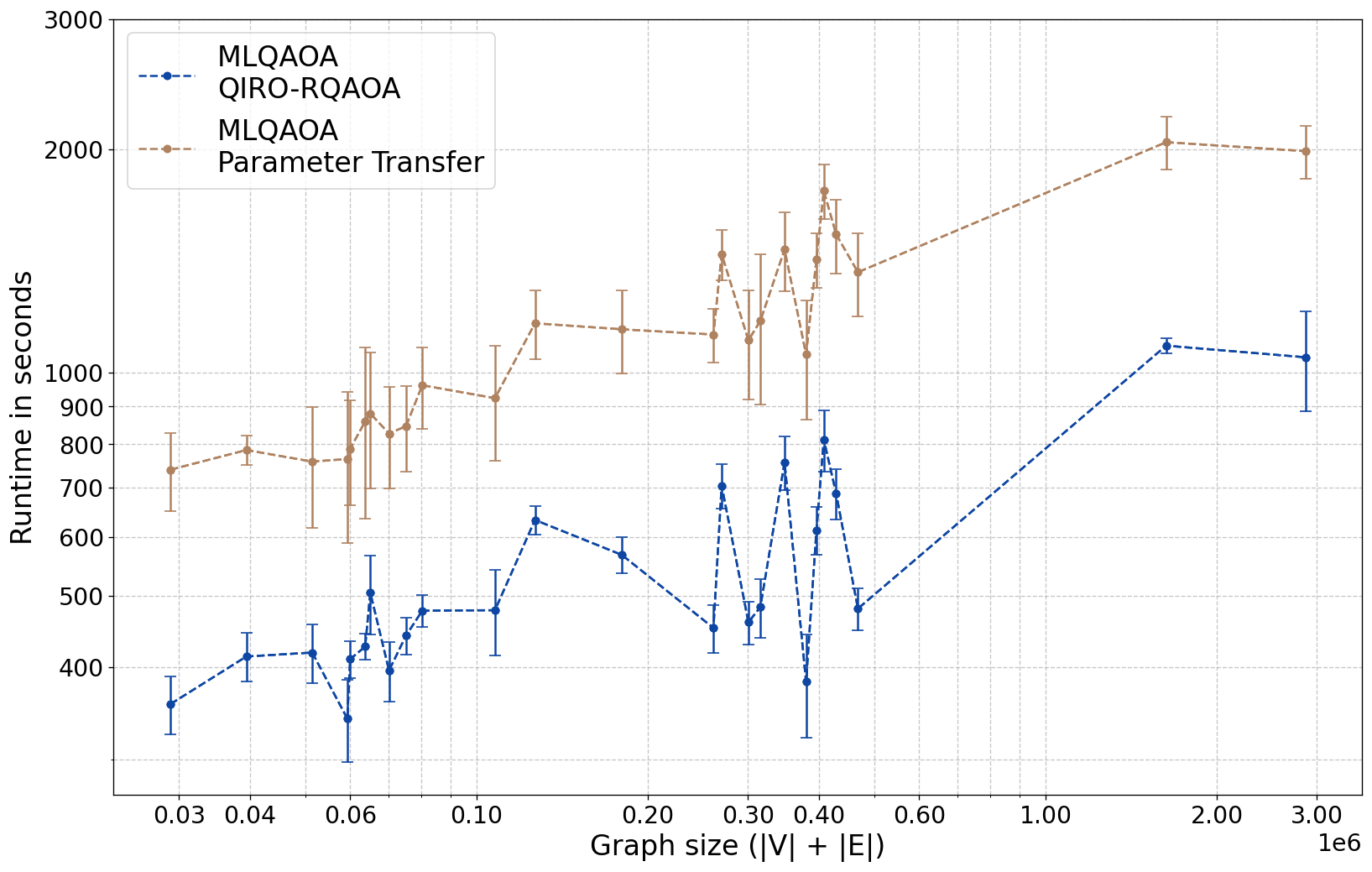}    
    \caption{Average run time in seconds of graph learning MLQAOA and RQAOA-inspired QIRO MLQAOA on graphs from table \ref{tab:total_table} over 20 runs. The $x-$axis and $y-$axis are logarithmic scales with $x$-axis denoted the runtime and $y-$axis denoted the size of the graph calculated by $\abs{V} + \abs{E}$.}
    \label{fig:runtime}
\end{figure*}

\paragraph{$G_{set}$ graphs} Table \ref{tab:Gset_table} shows a comparison between the proposed reinforced schemes MLQAOA and exhaustive learning multilevel MAXCUT QAOA \cite{angone2023hybrid} with respect to the optimal solution. In the first column, we show the details of graphs (number of vertices and edges) used from the $G_{set}$ dataset. In the second, third, and fourth columns, the approximation ratio of each method is recorded with the form: average approximation ratio / best approximation ratio. The final column shows the optimal cut (best objective value) for the MAXCUT problem for specific graphs. We observe that MLQAOA outperforms the exhaustive learning multilevel MAXCUT QAOA with a much shorter runtime. The running times of Graph Learning MLQAOA and RQAOA-QIRO MLQAOA are comparable (Fig. \ref{fig:runtime}), so while the quality of the latter is better, the gap with the former is not very significant.

\paragraph{Karloff graphs} We analyze the performance of MLQAOA against the Goemans-Williamson MAXCUT algorithm for the Karloff graphs in Table \ref{tab:karloff_graph}. The first column gives the details of Karloff graphs (number of vertices and edges), while the second column shows the average and best cuts from the Goemans-Williamson algorithm. The third and fourth columns demonstrate our average and best approximation ratio and the last column shows the optimal cut. The MLQAOA demonstrates competitive performance against the Goemans-Williamson algorithm for this type of graph, reaching the optimal cut in many cases.

\begin{figure}[!htb]
  \centering
  \includegraphics[width=0.49\textwidth, height=0.4\textwidth]{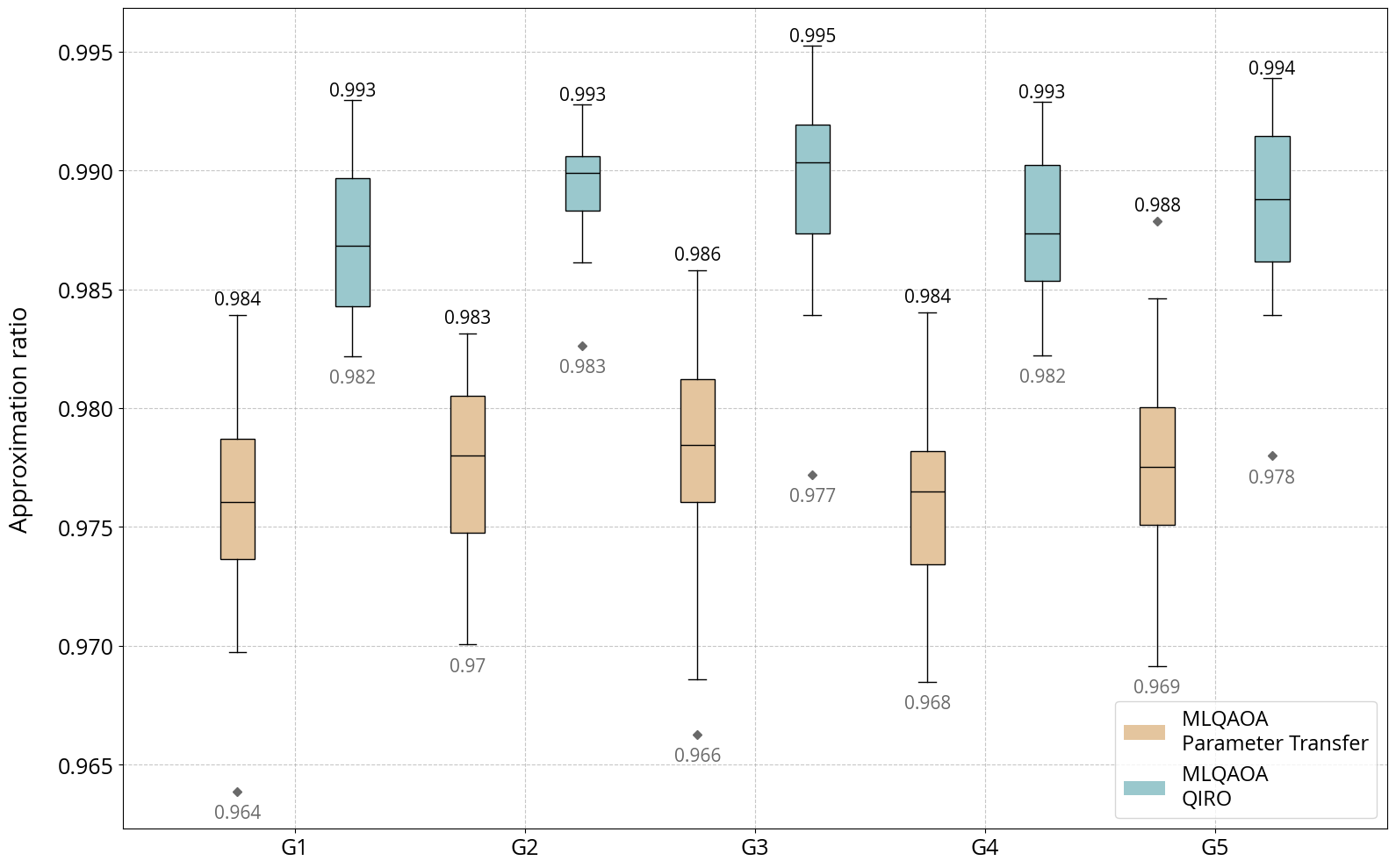}
  \hfill
  \includegraphics[width=0.49\textwidth, height=0.4\textwidth]{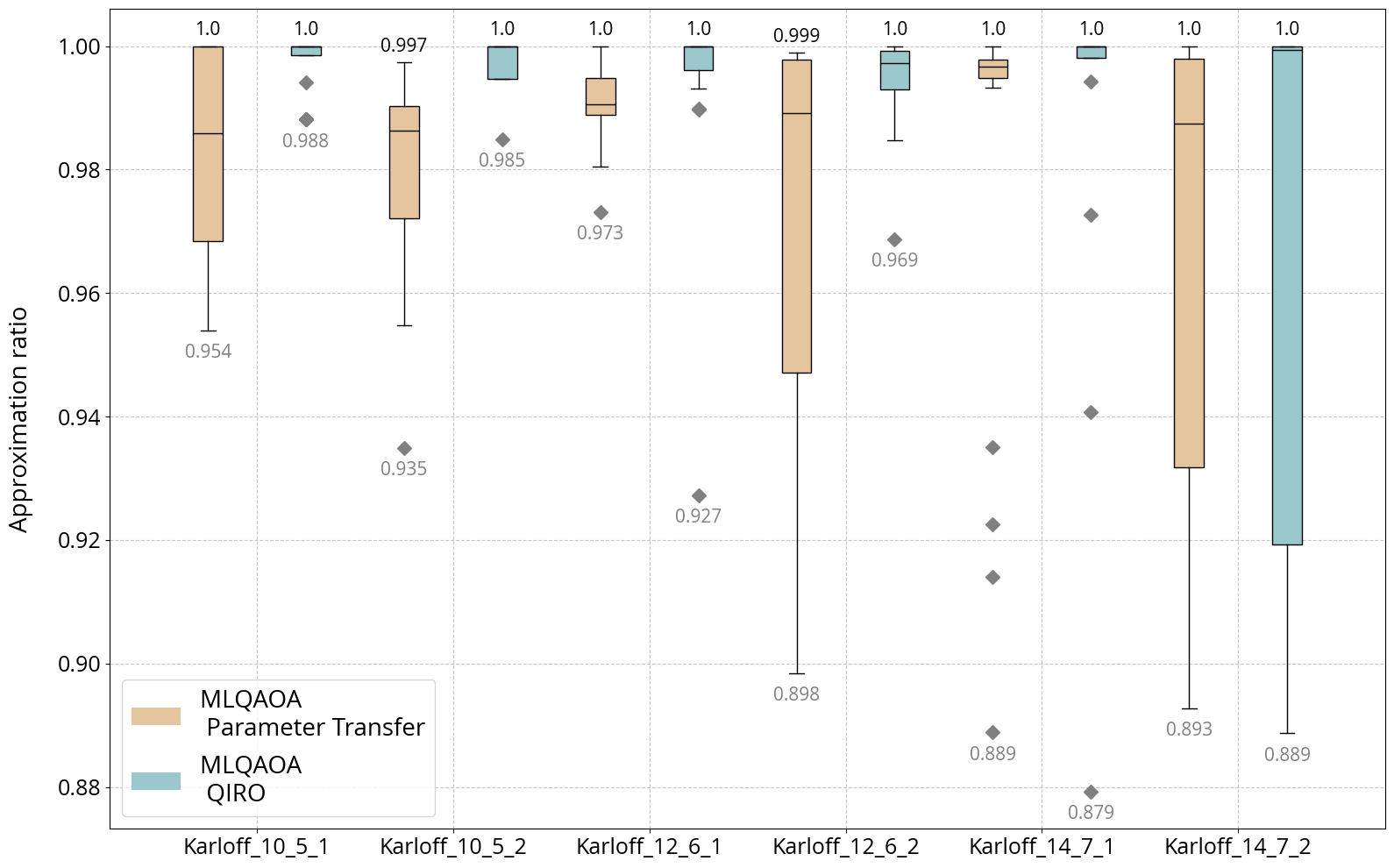}
  \caption{Approximation ratio of graph learning MLQAOA RQAOA-inspired QIRO MLQAOA on $G_{set}$ and Karloff graph.  The whisker lines are drawn up (down) to the largest (lowest) observed data point from the dataset that falls within the $1.5$ interquartile range (IQR) value from the upper (lower) quartile. The upper (lower) notated approximation ratio of each box indicates the highest (lowest) approximation ratio over $20$ run}
  \label{fig:Gset_Karloff_graphs}
\end{figure}

To provide a detailed insight into the performance of MLQAO on $G_{set}$ dataset and Karloff dataset, figure \ref{fig:Gset_Karloff_graphs} shows the boxplot of the graph learning MLQAOA and RQAOA-inspired QIRO MLQAOA. The boxes are drawn from the $25\%$ percentile to the $75\%$ percentile with a horizontal line drawn in the middle to denote the median. The $x$-axis represents the graph instances and the $y$-axis represents the approximation ratio. The RQAOA-inspired QIRO MLQAOA exhibits more consistent performance with less variance and achieves a slightly higher approximation ratio in most cases compared to graph-learning MLQAOA.

\paragraph{$G_{set}$ continued} We further compare our method with the QAOA-in-QAOA approach \cite{QAOA-in-QAOA} for solving more graphs from the $G_{set}$ dataset in table \ref{tab:extenede_Gset}. Our method demonstrates comparable performance with classical solvers like a dualscaling SDP solver (DSDP) \cite{DSDP} or a physics-inspired graph neural network (PI-GNN) method \cite{Schuetz2022},  outperforming the QAOA-in-QAOA approach but yielding worse results than breakout local search (BLS) \cite{BENLIC20131162}. These three graphs are separated from the previous graphs from $G_{set}$ due to the unavailability of the QAOA-in-QAOA code.

\paragraph{Larger graphs} Finally, to demonstrate the robustness of our method, we sample 25 graphs from the SuiteSparse Matrix Collection \cite{davis2011university} and the Network Repository \cite{Rossi_Ahmed_2015} and present their results in Table \ref{tab:total_table}. These graphs span different domains, sizes, and degrees, providing a comprehensive overview of our algorithm's performance compared to a systematic implementation of MAXCUT and QUBO heuristics in MQLib \cite{Dunning2018}. Among the classical heuristics, we choose the best performing to the MAXCUT problem including ``BURER2002'' \cite{BURER2002} (the best heuristic evaluated by different metrics from \cite{Dunning2018}), ``DUARTE2005'' \cite{DUARTE2005}, ``GLOVER2010'' \cite{glover2010diversification} and ``DESOUSA2013'' \cite{DESOUSA2013}. The reinforcement of MLQAOA gives a comparable result with classical solvers, especially for huge graphs as shown in table \ref{tab:total_table}. This holds a promise for hybrid-quantum algorithms to address other quadratic unconstrained binary optimization problems (QUBO) such as Maximum independent set or max-$k$ satisfiability.

To conclude the experiments and numerical simulation, Figure \ref{fig:runtime} illustrates the runtime of our reinforced method for various graph sizes. This log-scale graph highlights the scalability and potential of our method when executed on quantum hardware. The execution time shown here is from simulation and can be significantly reduced when implemented on quantum devices.

\begin{table}[]
    \centering
    \caption{$G_{set}$ graphs approximation ratio between three different methods: exhaustive learning MLQAOA, graph-learning MLQAOA, and RQAOA-inspired QIRO MLQAOA. The approximation ratio is recorded over $20$ runs. The exhaustive MLQAOA \cite{angone2023hybrid} has not been executed multiple times due to the incomparable slowness.}
    \label{tab:Gset_table}
    \begin{tabular}{l|l|l|l|l}
    \hline
    $G(\abs{V},\abs{E})$ &   MLQAOA \cite{angone2023hybrid}  &  \begin{tabular}[c]{@{}l@{}} MLQAOA\\ Graph-Learning\end{tabular} &  \begin{tabular}[c]{@{}l@{}} MLQAOA\\ RQAOA-QIRO\end{tabular} &  Optimal Cut\\ \hline
    $G_1(800, 19716)$ & -/0.984                                                               & 0.976/0.985                                                             & \textbf{0.989/0.993} & 11624                                            \\ 
    $G_2(800, 19716)$ & -/0.984                                                                & 0.978/0.984                                                              & \textbf{0.989/0.993} & 11620                                             \\ 
    $G_3(800,  19716)$ & -/0.982                                                                & 0.978/0.986                                                              & \textbf{0.990/0.995}  & 11622                                           \\ 
    $G_4(800,  19716)$ & -/0.978                                                                & 0.980/0.988                                                              & \textbf{0.990/0.994} & 11646                                             \\ 
    $G_5(800,  19716)$ & -/0.983                                                                & 0.979/0.989                                                             & \textbf{0.989/0.994} & 11631                                             \\
    \hline
\end{tabular}
\end{table}

\begin{table}[]
    \centering
    \caption{Karloff graphs approximation ratio between three different methods: Goemans-Williamson MAX CUT algorithm, graph learning MLQAOA, and RQAOA-inspired QIRO MLQAOA. The approximation ratio is recorded over $20$ runs}
    \label{tab:karloff_graph}
\begin{tabular}{l|l|l|l|l}
\hline
&\textbf{Classical} & \multicolumn{2}{c}{\textbf{Quantum approach}} & \\ \hline
$G(\abs{V}, \abs{E})$   & GW \cite{GW_algorithm}      & \begin{tabular}[c]{@{}l@{}}MLQAOA\\ Graph-Learning\end{tabular} & \begin{tabular}[c]{@{}l@{}}MLQAOA\\ RQAOA-QIRO\end{tabular}  &  Optimal Cut\\ \hline
$K(252, 3150)$ & 0.881/0.925 & 0.984/{\bf 1.0}                                                              & \textbf{0.997/1.0}  & 2520                                                             \\ 
$K(252, 12600)$ & 0.940/0.941 & 0.981/0.997                                                            & \textbf{0.997/1.0}  & 7560                                                           \\ 
$K(924, 16632)$ & 0.879/0.913 & 0.991/{\bf 1.0}                                                            & \textbf{0.994/1.0}  & 13860                                                               \\ 
$K(924, 103950)$ & 0.912/0.922  & 0.988/0.999                                                              & \textbf{0.995/1.0} & 69300                                                               \\
$K(3432, 84084)$ & 0.879/0.937 & \textbf{0.992/1.0}                                                             & 0.989/{\bf 1.0} & 72072                                                  \\ 
$K(3432, 756756)$ & 0.897/0.927 & \textbf{0.969/1.0}                                                              & 0.968/{\bf 1.0}  & 540540                                                         \\ \hline
\end{tabular}
\end{table}

\begin{table*}[t]
\centering
\caption{Extended $G_{set}$ graphs approximation ratio between different quantum and classical methods, including graph-learning MLQAOA, RQAOA-inspired QIRO MLQAOA, and QAOA-in-QAOA. The best approximation ratio for both classical and quantum algorithms is recorded. The best results in their categories are in bold.}
\label{tab:extenede_Gset}
\small 
\begin{tabular}{l|l|l|l|l|l|l|l|l|l}
\hline
\multicolumn{3}{c|}{} & \multicolumn{4}{|c|}{\textbf{Quantum approach}} & \multicolumn{3}{|c}{\textbf{Classical approach}} \\ \hline
\textbf{Graph} & $\abs{V}$ & $\abs{E}$ & \begin{tabular}[c]{@{}l@{}}MLQAOA\\ Graph Learning\end{tabular} & \begin{tabular}[c]{@{}l@{}}MLQAOA\\ RQAOA-QIRO\end{tabular} & \begin{tabular}[c]{@{}l@{}}QAOA$^{2}$\\ $p = 1$\end{tabular} & \begin{tabular}[c]{@{}l@{}}QAOA$^{2}$\\ $p = 4$ \end{tabular} & DSDP \cite{DSDP} & PI GNN \cite{Schuetz2022} & BLS \cite{BENLIC20131162}  \\ \hline
$G_{14}$   & 800   & 4694  & 2994  & {\bf 3026}  & 2593                                                                        & 2596                                                                        & 2922                      & 3026                      & \textbf{3064}  \\ 
$G_{15}$   & 800   & 4694  & 2977                                                            & {\bf 3026}                                                        & 2596                                                                        & 2579                                                                        & 2938                      & 2990                      & \textbf{3050}  \\ 
$G_{22}$   & 2000  & 19990 & 13122                                                           & {\bf 13174}                                                       & 10664                                                                       & 10559                                                                       & 12960                     & 13181                     & \textbf{13359} \\ \hline
\end{tabular}
\end{table*}

\begin{table*}[t]
\centering
\caption{Best objective cut of graph-learning MLQAOA (MLQAOA GL) and RQAOA-inspired QIRO (MLQAOA QIRO) compared with classical heuristic algorithm ``BURER2002'' \cite{BURER2002}, ``DUARTE2005'' \cite{DUARTE2005}, ``GLOVER2010'' \cite{glover2010diversification} and ``DESOUSA2013'' \cite{DESOUSA2013}. The graphs are drawn from SuiteSparse Matrix Collection \cite{davis2011university} and the Network Repository\cite{Rossi_Ahmed_2015}.} 
\label{tab:total_table}
\small 
\begin{tabular}{|l|l|l|l|l|l|l|l|l|}
\hline
\multicolumn{3}{|c|}{} &  \multicolumn{2}{|c|}{\textbf{Quantum approaches}} & \multicolumn{4}{|c|}{\textbf{Classical approaches}}\\ \hline
\small \textbf{Graph} & $\abs{V}$ & $\abs{E}$    & \small \begin{tabular}[c]{@{}l@{}} MLQAOA\\ GL\end{tabular} &\small \begin{tabular}[c]{@{}l@{}}MLQAOA\\ QIRO\end{tabular} & \small BURER02 & \small DUARTE05 & \small GLOVER10  & \small DESOUSA13 \\ \hline
soc-buzznet & 101,163 &  2,763,066     & 2,069,342                                                          & 2,070,569                                                   & 2,071,445 & \textbf{2,071,612}  & 2,071,427   & 1,390,645\\ \hline %\midrule
c-72 & 84,064  &  395,811         & 311,299    &\textbf{311,623}      & 296,238  & 303,000   & 289,878    & 158,740\\ \hline 
c-71 & 76,638  &  468,096        & 390,832        & \textbf{391,333}      & 391,007  & 390,937   & 360,192    & 198,915\\ \hline
soc-slashdot & 70,068  &  358,647     & 277,664      & 278,124       & 277,675  & \textbf{278,445}   & 276,640 & 182,254   \\ \hline
c-68 & 64,810  &  315,408      & 250,405        & \textbf{250,547}    & 242,676  & 245,123   & 233,846  & 127,692  \\ \hline
dixmaanl & 60,000  &  179,999        & 96,299         & 96,659              & \textbf{99,555}   & 98,281    & 89,411   & 61,713  \\ \hline
soc-brightkite & 56,739  &  212,945   & 149,739          & 150,183          & 149,328  & \textbf{151,057}   & 145,405 & 108,795  \\ \hline
copter2 & 55,476  &  407,714         & 218,486      & 219,595        & \textbf{225,982}  & 222,862   & 213,634 & 178,894  \\ \hline
c-64 & 51,035  &  384,438             & 328,421    & 328,471         & \textbf{328,475}  & \textbf{328,475}   & 326,302  & 169,408  \\ \hline
3dtube & 45,330  &  1,629,474          & 1,033,982      & 1,044,174      & 1,065,693 & \textbf{1,066,489}  & \textbf{1,066,489}  & 797,620  \\ \hline
c-62 & 41,731  &  300,537           & 258,698          & \textbf{258,805}      & 258,759  & 258,802   & 245,026   & 132,373 \\ \hline
c-59 & 41,282  &  260,909           & 219,481       & \textbf{219,618}    & 218,368  & 219,193   & 216,117  & 112,342  \\ \hline
shock-9 & 36,476  &  71,290         & 66,702 & 69,168    & \textbf{69,772}   & 68,717    & 64,780 & 37,034    \\ \hline
big\_dual & 30,269  &  44,929       & 41,658                                                             & 41,933                                                     & \textbf{43,280}   & 42,489    & 39,801  & 23,551   \\ \hline
rajat10 & 30,202  &  80,202         & 38,776                                                             & 38,989                                                      & \textbf{39,847}   & 39,316    & 36,028 & 26,223    \\ \hline
aug2dc & 30,200  &  40,000          & 38,221                                                             & 38,407                                                      & \textbf{39,805}   & 38,670    & 31,920 & 21,080    \\ \hline
soc-epinions & 26,588  &  100,120     & 68,934                                                             & 69,282                                                    & 69,496   & \textbf{69,850}    & 67,541 & 51,705    \\ \hline
dtoc & 24,993  &  34,986            & 33,915                                                             & 34,070                                                      & \textbf{34,590}   & 33,951    & 32,454  & 18,497   \\ \hline
rajat09 & 24,482  &  64,982        & 31,514                                                             & 31,588                                                      & \textbf{32,334}   & 31,874    & 29,136 & 21,305    \\ \hline
aug3d & 24,300  &  34,992           & 34,784                                                             & 34,785                                                     & \textbf{34,992}   & 33,866    & 27,590  & 18,462   \\ \hline
biplane-9 & 21,701  &  42,038       & 39,740                                                             & 39,892                                                     & \textbf{41,373}   & 40,660    & 41,120 & 22,129    \\ \hline
rajat08 & 19,362  &  51,362         & 24,913                                                             & 24,972                                                     & \textbf{25,581}   & 25,183    & 23,228  & 16,968   \\ \hline
ex3sta1 & 16,782  &  34,7890        & 180,530                                                            & 181,141                                                    & \textbf{183,535}  & 183,141   & 180,975  & 168,116  \\ \hline
rajat07 & 14,842  &  39,342       & 19,078                                                             & 19,144                                                      & \textbf{19,575}   & 19,303    & 17,749   & 13,115  \\ \hline
rajat06 & 10,922  &  28,922        & 14,034                                                             & 14,083                                                     & \textbf{14,418}   & 14,211    & 13,203   & 9732   \\ \hline
\end{tabular}
\end{table*}

\section{Discussion}
%This study showcases the integration of quantum algorithms as sub-problem solvers within the framework of Multilevel MAXCUTQAOA \cite{angone2023hybrid} and gives a comparative analysis across various solvers. In more detail, through the utilization of graph learning techniques on weighted graphs for parameter transfer and the adoption of RQAOA-inspired QIRO, we address the weakness of excessive runtime of Multilevel MAXCUT QAOA. This strategic enhancement significantly reduces the overall runtime, thereby facilitating experiments with larger graph problems.
%The improved runtime permits multiple iterations on the same problem instance to attain the best objective. This iterative approach mirrors the spirit of classical multilevel algorithms, leading to substantial improvement in the final objective. Furthermore, our comprehensive analysis reveals that MLQAOA exhibits notably robust performance, particularly on denser graphs that traditionally pose challenges for well-known classical algorithms. This observation underscores the potential of our quantum algorithms scheme in addressing large-scale problems, offering promising advantages, especially with the constraints of the NISQ era.

%An interesting future research direction that we believe can further enhance the MLQAOA scheme is a more quantum native coarsening phase and a better approach to choosing sub-problem graphs.

The main goal successfully achieved in this work was to break the variational quantum algorithm complexity barriers to tackle large-scale combinatorial optimization instances without losing the solution quality. As one can see from the results both Graph Learning MLQAOA and RQAOA-QIRO MLQAOA are highly scalable, and pretty similar to each other in terms of the solution quality (while RQAOA-QIRO MLQAOA is usually slightly better). They both often reach optimality (if it is known for the comparison) and are identical to the MAXCUT dedicated top heuristics. It is important to mention that no optimization of hyperparameters in this work has been done and it is clear that these results could be further improved. At the same time, we note that no comparison with generic solvers (such as Gurobi) has been presented because they are either way slower than MLQAOA or exhibit low quality. Below we discuss several important lessons learned and obstacles we encountered during this work.

\paragraph{Graph representation learning} One unexpected result was a relatively low quality of the graph representation learning based on the Weisfeiler-Leman (W-L) graph isomorphism test (e.g., graph2vec algorithm) in comparison to the spectral learning technique. In the previous work \cite{jose2024graph},  the QAOA parameter transferability based on the W-L test was more successful on the unweighted graphs than anything else. Although we observe the high quality and scalability of the spectral representation learning transferability, more investigation is required to build a weighted graph transferability model.

\paragraph{Coarsening scheme} The current multilevel scheme was significantly simplified to understand the quantum-based refinement effects. However, similar to many multilevel solvers for graphs, this can be improved. Introducing advanced algebraic multigrid coarsening \cite{safro2012advanced,brandt2003multigrid} opens an opportunity to preserve the spectral properties of the original problem at all levels of coarseness much better. As a result, potentially fewer refinement steps are anticipated.

\paragraph{RQAOA-QIRO} Using the idea of recursive QAOA for simplification rule leads to the QIRO scheme demonstrating excellent performance on  MAXCUT. The QIRO was chosen as a baseline because it was consistently outperforming its competitors on small-scale graphs. However, similar to many other methods with the goal of creating a more compact circuit, this method encounters scalability challenges in terms of both time and space complexity when applied to large instance graphs. An interesting observation emerges regarding the RQAOA-QIRO approach, the approach produces consistent results for large graphs, indicating its tendency to converge to local optima. Conversely, the graph learning approach yields more fluctuated results with a higher degree of randomness, leading to a better chance of escaping local points and heading toward global minimum. On average, the difference between graph learning MLQAOA and RQAOA-QIRO MLQAOA is only 0.7\% on the set of larger graphs, i.e., they are both highly competitive.
% Nevertheless, combined with the multilevel scheme, this approach gives a comparable result with classical solvers, especially for huge graphs as shown in table \ref{tab:total_table}. This holds a promise for hybrid-quantum algorithms to address quadratic unconstrained binary optimization problems (QUBO) such as Maximum independent set or max-$k$ satisfiability.

% Incorporating quantum algorithms as sub-problem solvers yields the potential to bring in quantum advantage in the optimization fields. In the work, the coarsening phase still depends on classical computation. Therefore, to make the coarsening level native to quantum information, an interesting approach is to treat the original graph as Ising Hamiltonian and well-studied methods in the field of Physics to coarsen the Hamiltonian such as the renormalization group. Moreover, there is still room to improve things such as the interpolation of information from the previous level to the current level or how to choose sub-problems such that their solution contributes the most to the current level solution. Another nice thing to consider is how the construction of sub-problems and coarsening relate to the actual implementation constraints of quantum computers. By devising sub-problem graphs that isomorphism to quantum device architecture, we can avoid $SWAP$ gates overhead and increase the credibility of the solution. 

\section*{Acknowledgment}

This work was supported with funding from the Defense Advanced Research Projects Agency (DARPA) under the ONISQ program.

\bibliographystyle{unsrt}
\bibliography{bibliography}

\end{document}